\documentclass[floatfix,lengthcheck,showpacs,amssymb,amsmath,amsfonts,twocolumn]{aastex63}
\usepackage{CJK}
\usepackage{lineno}
\usepackage[utf8]{inputenc}
\usepackage{color}
\usepackage{graphicx}
\usepackage{float}
\usepackage{subfigure}
\usepackage{amsmath}
\usepackage{afterpage}
\usepackage{acronym}
\usepackage{multirow}
\usepackage{tikz}
\usetikzlibrary{arrows,shapes,trees,decorations.pathreplacing}
\usepackage{import}
\usepackage{svn-multi}
\usepackage{hyperref}
\usepackage{gensymb}
\usepackage{longtable}
\hypersetup{
    colorlinks=true,
    linkcolor=blue,
    filecolor=magenta,      
    urlcolor=cyan,
}

\newcommand{\msun}{\ensuremath{\mathrm{M}_{\odot}}}

\usepackage{xcolor,colortbl}
\definecolor{o1}{gray}{0.9}
\definecolor{o2}{gray}{0.8}
\definecolor{o3}{gray}{0.6}
\definecolor{Gray}{gray}{0.9}
\usepackage{longtable}

\begin{document}
\begin{CJK*}{UTF8}{gbsn}
\title[]{3-OGC: Catalog of gravitational waves from compact-binary mergers}

\correspondingauthor{Alexander H. Nitz}
\email{alex.nitz@aei.mpg.de}

\author[0000-0002-1850-4587]{Alexander H. Nitz}
\author[0000-0002-0355-5998]{Collin D. Capano}
\author[0000-0002-6404-0517]{Sumit Kumar}
\author[0000-0002-2928-2916]{Yi-Fan Wang (王一帆)}
\author[0000-0003-0966-1748]{Shilpa Kastha}
\author[0000-0002-6990-0627]{Marlin Sch{\"a}fer}
\author[0000-0002-5077-8916]{Rahul Dhurkunde}
\affil{Max-Planck-Institut f{\"u}r Gravitationsphysik (Albert-Einstein-Institut), D-30167 Hannover, Germany}
\affil{Leibniz Universit{\"a}t Hannover, D-30167 Hannover, Germany}
\author[0000-0003-4059-4512]{Miriam Cabero}
\affil{Department of Physics and Astronomy, The University of British Columbia, Vancouver, BC V6T 1Z4, Canada,}


\begin{abstract}
We present the third Open Gravitational-wave Catalog (3-OGC) of compact-binary coalescences, based on the analysis of the public LIGO and Virgo data from 2015 through 2019 (O1, O2, O3a). Our updated catalog includes a population of 57 observations, including four binary black hole mergers that had not previously been reported. This consists of 55 binary black hole mergers and the two binary neutron star mergers GW170817 and GW190425. We find no additional significant binary neutron star or neutron star--black hole merger events. The most confident new detection is the binary black hole merger GW190925\_232845 which was observed by the LIGO Hanford and Virgo observatories with $\mathcal{P}_{\textrm{astro}} > 0.99$; its primary and secondary component masses are $20.2^{+3.9}_{-2.5} M_{\odot}$ and $15.6^{+2.1}_{-2.6} M_{\odot}$, respectively. We estimate the parameters of all binary black hole events using an up-to-date waveform model that includes both sub-dominant harmonics and precession effects. To enable deep follow-up as our understanding of the underlying populations evolves, we make available our comprehensive catalog of events, including the sub-threshold population of candidates, and the posterior samples of our source parameter estimates.
\end{abstract}

\section{Introduction}

With the advent of the current generation of interferometric gravitational-wave detectors, the observation of gravitational waves from the coalescence of compact-binary mergers has become a regular and rapidly maturing component of astronomy. The Advanced LIGO~\citep{TheLIGOScientific:2014jea} and Advanced Virgo~\citep{TheVirgo:2014hva} observatories have now been observing at high sensitivity since 2015 and 2017, respectively. During this period they have completed three observing runs (O1-O3). Dozens of binary black hole mergers have been reported from these observing runs, in addition to a handful of binary neutron star coalescences~\citep{Nitz:2019hdf, Venumadhav:2019lyq, LIGOScientific:2018mvr,Abbott:2020niy,Abbott:2020uma, TheLIGOScientific:2017qsa}. Notably, GW170817 remains the sole observation with unambiguous electromagnetic counterparts~\citep{GBM:2017lvd,TheLIGOScientific:2017qsa}. Novel observations such as the massive GW190521 merger~\citep{Abbott:2020tfl} are starting to challenge our models of stellar formation~\citep{Abbott:2020mjq,Gerosa:2021mno,Edelman:2021fik,Zevin:2020gbd} and are pushing the limits of gravitational waveform modelling~\citep{Romero-Shaw:2020thy,Gayathri:2020coq,Estelles:2021jnz}.

\begin{figure*}[t]
  \centering
    \hspace*{-1.8cm}\includegraphics[width=2.6\columnwidth]{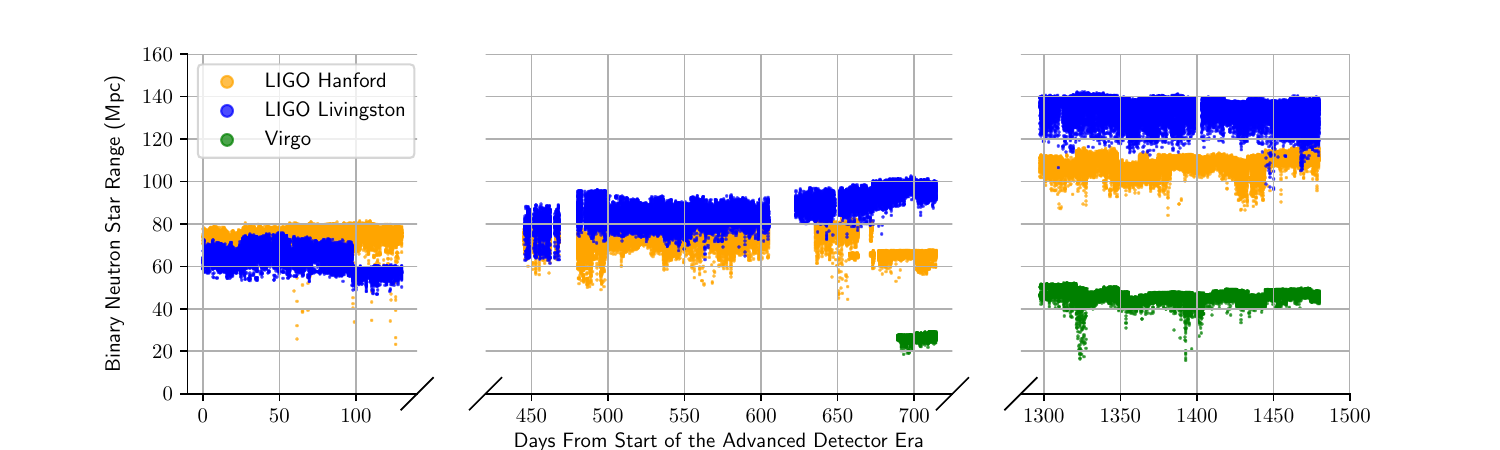}
\caption{The sky and orientation averaged distance that a fiducial 1.4-1.4 \msun BNS merger can be observed by the LIGO Hanford (yellow), LIGO Livingston (blue) and Virgo (green) observatories at an SNR of 8. The O1 (left), O2 (middle) and O3a (right) observing periods are shown.}
\label{fig:range}
\end{figure*}

In this work, we provide a comprehensive catalog of gravitational waves from the coalescence of binary neutron star (BNS), neutron star--black hole (NSBH) and binary black hole (BBH) systems based on a deep archival search for compact-binary mergers of the public LIGO and Virgo data~\citep{Vallisneri:2014vxa,Abbott:2019ebz}. The previous open gravitational-wave catalog (2-OGC) searched for the signature of compact-binary mergers in the O1 and O2 observing runs. We re-analyze the entirety of the public LIGO and Virgo data comprised of O1, O2, and the recently published O3a dataset~\citep{Vallisneri:2014vxa,Abbott:2019ebz}, which covers the first half, from April 1 to October 1 of 2019, of the concluded O3 observing run. The O3 data is being released in 6-month chunks, with O3a being the first; the second half is expected in 6 months time. Included in our data release is the complete set of sub-threshold candidates in addition to posterior samples from estimates of the most significant mergers. Sub-threshold candidates can be correlated with archival observations (e.g. from gamma-ray bursts~\citep{Burns:2018pcl,Nitz:2019bxt}, high-energy neutrinos~\citep{Countryman:2019pqq}, or optical transients~\citep{Andreoni:2018fcm,Setzer:2018ppg}) to potentially uncover fainter, distant populations.

We improve the sensitivity of our analysis over our previous catalog search by targeted use of signal-consistency tests, updated data cleaning procedures, and stricter allowance for loss in signal-to-noise. As in 2-OGC, for candidates consistent with the bulk of the increasing population of observed BBH mergers, we estimate the probability of astrophysical origin using the focused BBH region of our larger search. This estimate takes into account the measured rate of mergers and the possibly confounding background noise. Additionally, for the first time in this catalog, we incorporate BNS and BBH candidates observed by a single sensitive detector using methods introduced in~\cite{Nitz:2020naa}. 

We find that 55 binary black hole mergers have been observed from 2015-2019 along with 2 binary neutron star mergers. These include four BBH mergers from the O3a period which had not previously been reported. Our results are broadly consistent with the cumulative sum of previous catalogs~\citep{Nitz:2018imz, Nitz:2019hdf, Venumadhav:2019lyq, LIGOScientific:2018mvr}, including the recent analysis of O3a by the LVK collaboration~\citep{Abbott:2020niy}.

\begin{table}[b]
  \begin{center}
    \caption{Analyzed time in days for different instrument observing combinations. We use here the abbreviations H, L, and V for the LIGO-Hanford, LIGO-Livingston, and
    Virgo observatories respectively. Only the indicated combination of observatories were operating for each time period, hence each is exclusive of all others. Some data $O(1)\%$ is excluded due to analysis requirements.}
    \label{table:data}
\begin{tabular}{rrrrrrrr}
Observation & HLV & HL & HV & LV & H & L & V \\\hline
O1 \vline & - & 48.6 & - & - & 27.6 & 17.0 & - \\
O2 \vline & 15.2 & 103.3 & 1.7 & 2.2 & 37.8 & 33.0 & 1.7 \\
O3a \vline & 79.7 & 26.1 & 17.4 & 25.2 & 5.6 & 6.4 & 15.7 \\
All \vline & 95.0 & 178.0 & 19.1 & 27.4 & 70.9 & 56.4 & 17.4 \\
\end{tabular}
  \end{center}
\end{table}

\section{LIGO and Virgo Observing Period}

We analyze the complete set of public LIGO and Virgo data from the O1, O2, and O3a observing runs~\citep{Vallisneri:2014vxa,Abbott:2019ebz}.  In our analysis, we also include data around GW170608~\citep{Abbott:2017gyy} and GW190814~\citep{Abbott:2020khf} which were released separately~\citep{Vallisneri:2014vxa,Abbott:2019ebz}. The data sets have been calibrated by the LIGO Scientific and Virgo Collaborations to convert the optical signals at the readout ports of the interferometers into timeseries of dimensionless strain using photon calibrator systems as length fiducials~\citep{Viets:2017yvy, Acernese:2018bfl, Bhattacharjee:2020yxe, Estevez:2020pvj}. Additionally, the LIGO and Virgo datasets have undergone noise subtraction to remove persistent noise sources measured using witness auxiliary sensors~\citep{Davis:2018yrz, Vajente:2019ycy, Estevez2019, Rolland2019}. Finally, data quality categories based on information of the detectors and investigations of noise sources during the observing run are provided to reduce the number of false alarms~\citep{Davis:2021ecd}.

The time evolution of the BNS range for each observatory and the distribution of detector observing times are shown in Fig.~\ref{fig:range} and Table~\ref{table:data}, respectively. 
In total, there have been $464$ days of Advanced LIGO and Virgo observing time. Two or more detectors were observing during $320$ days, of which $95$ days were joint observations of the full LIGO-Hanford, LIGO-Livingston and Virgo network. 

The newly released data from O3a adds 176 days of observational data to the existing 288 days from O1 and O2. During the period of O3a, $79.7$ days have data from all 3 detectors available, $68.7$ days rely on a 2 detector network and $27.7$ days contain data only from a single detector. Several upgrades were implemented at the LIGO and Virgo detectors between O2 and O3a to improve the sensitivity of the detectors~\citep{Abbott:2020niy, Buikema:2020dlj}. The maximum BNS range throughout O3a was 142.4 Mpc for LIGO-Livingston, 117.2 Mpc for LIGO-Hanford, and 52.2 Mpc for Virgo.

For the first time, we also include candidates occurring during the 174 days when only one single Advanced LIGO detector was observing. We do not include single-detector candidates from the 17.4 days of Advanced Virgo data.

\section{Search for Compact-binary Mergers}

 We use matched filtering to extract the signal-to-noise ratio (SNR) of a potential signal~\citep{Allen:2005fk,findchirp}, as is the standard procedure for the most sensitive gravitational-wave searches where there is an accurate model of the gravitational waveform available~\citep{Davies:2020,gstlal-methods,Venumadhav:2019tad}. We assess each potential candidate for consistency with the expected gravitational-wave morphology~\citep{allen:2004gu, Nitz:2017lco} and then rank potential candidates~\citep{Davies:2020,Mozzon:2020gwa} based on factors including the overall noise rate and each signal's coherence between detectors~\citep{Nitz:2017svb}. We require a minimum SNR of 4 from each detector which contributes to a candidate.
 
 The procedure broadly follows the same methods used to construct the prior 2-OGC catalog~\citep{Nitz:2019hdf}, but with improvements to the removal of loud transient glitches and more stringent constraints on our suite of signal consistency tests. Detailed configuration files necessary to reproduce the analysis are included in our data release~\citep{3-OGC}. In addition, we use a denser bank of templates to reduce loss in sensitivity from mismatch between our template bank and the gravitational-wave signal. The analysis is accomplished using the public and open source PyCBC analysis toolkit~\citep{pycbc-github}.

\begin{figure}[t]
  \centering
    \includegraphics[width=\columnwidth]{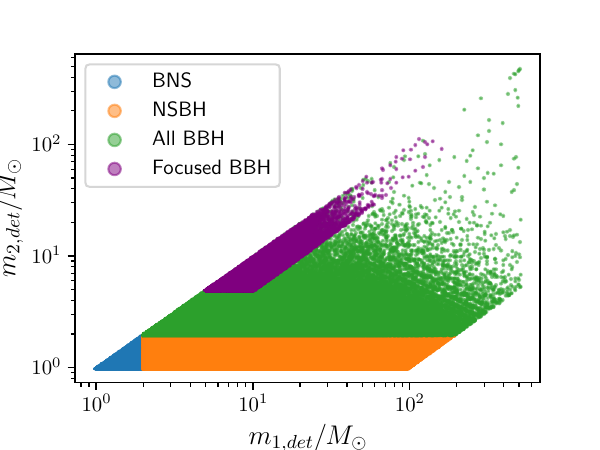}
\caption{The detector-frame (redshifted) component masses of templates used to search for compact binary mergers. The template bank is constructed in four parts using stochastic placement. The binary neutron star (blue), neutron star--black hole (orange), binary black hole (green), and focused binary black hole (purple) regions are shown. Templates in the focused binary black hole region are colored in purple and bounded by $m_{1,2} > 5\,\msun$,  $m_1/m_2 < 2$, and $m_{1, det} + m_{2,det} <250\,\msun$. The detector-frame masses are related to the source-frame masses through the redshift (z) by $m_{1/2,det} = m_{1/2}(1 + z)$, which accounts for the effects of cosmic expansion. This region is responsible for the vast majority of observed BBH mergers, but is composed of only $\sim3\%$ of the total number of templates ($\sim 1.3$ million), despite placement at higher density than the rest of the search space.
}
\label{fig:bank}
\end{figure}
\vspace{20pt}
\subsection{Search Space}

To search for gravitational-wave sources using matched filtering, we rely upon accurate models of the gravitational waveform to act as templates. To account for sources with varied component masses and component spins, we construct a discrete bank of templates designed to ensure that for any signal within the target region there is a matching template able to recover its SNR at a prescribed maximum loss. We note that a different criteria would be used to maximize detections at a fixed computational cost~\citep{Allen:2021yuy}, however, this analysis is not computational limited. As shown in Fig.~\ref{fig:bank}, the search region can be divided into three parts targeting different types of sources, namely, BNS (blue), NSBH (orange), and BBH (green and purple) sources. The template bank is designed to detect non-precessing sources in quasi-circular orbits which can be modelled by two component masses and the spin of each component parallel to the orbital angular momentum. For BNS sources, we allow for matter effects up to $\tilde{\Lambda} < 300$, where $\tilde{\Lambda}$ is a weighted average of the component stars' tidal deformabilities~\citep{Flanagan:2007ix}. Inclusion of these effects may become important for future detectors, though have only minor impacts on the sensitivity of current searches; we include the effect here as a proof of principle~\citep{Harry:2021hls}.

The broad BNS, NSBH, and BBH regions are constructed so that signals lose no more than $3\%$ in SNR due to the discreteness of the template bank. The boundaries are similar to those used in our previous catalog~\citep{DalCanton:2017ala}, however, we no longer restrict each template's duration and instead include templates with component masses up to $500\msun$ (detector frame). In addition, there is a separate focused BBH region (shown in purple in Fig.~\ref{fig:bank}) which contains the entirety of known binary black hole sources with the exception of the high mass-ratio merger GW190814~\citep{Abbott:2020khf}. To ensure the maximum sensitivity to faint signals, we place templates in this region ensuring no more than $0.5\%$ of SNR is lost due to bank discreteness. Stochastic placement~\citep{Harry:2009ea,Ajith:2012mn} as implemented in the PyCBC toolkit~\citep{pycbc-github} is used to construct each bank. 

Despite targeting non-precessing sources, we expect this search retains sensitivity to some types of moderately precessing sources~\citep{TheLIGOScientific:2016qqj}, especially if they are short in duration or have orientation near to face-on/off. Searches which neglect precession lose sensitivity to highly precessing sources if the sources are a combination of high mass ratio, highly inclined, and observable for many cycles~\citep{Harry:2016ijz}. Similarly, we expect this search to lose sensitivity to highly eccentric sources~\citep{Ramos-Buades:2020eju, Wang:2021qsu}. Separate searches have been conducted focusing on eccentric sources~\citep{Nitz:2021mzz,Nitz:2019spj,Salemi:2019owp} and on those sources outside of the regions we consider here~\citep{Nitz:2020bdb,Authors:2019qbw}. Where the search methodology or waveform modelling is not yet sufficient, alternate techniques based on looking for coherent excess power are employed on LIGO and Virgo data~\citep{Klimenko:2008fu,Klimenko:2015ypf,Tiwari:2015gal}.

We employ three waveform models within our search: TaylorF2~\citep{Sathyaprakash:1991mt,Droz:1999qx,Blanchet:2002av,Faye:2012we}, IMRPhenomD~\citep{Husa:2015iqa,Khan:2015jqa}, and the reduced order model of SEOBNRv4~\citep{Taracchini:2013,Bohe:2016gbl} as implemented in~\citep{lalsuite}. TaylorF2 models only the inspiral portion of a gravitational-wave signal and is suitable for cases where the merger would be hidden by the detector noise. As such, it is employed only in the BNS region of our analysis. Where applicable, TaylorF2 includes tidal corrections up to 7.0 post-Newtonian order~\citep{Vines:2011ud,Damour:2012yf}. IMRPhenomD is used within the focused BBH search (purple) and SEOBNRv4 is used everywhere else. Both IMRPhenomD and SEOBNRv4 model the inspiral, merger, and ringdown of a non-precessing binary black hole coalescence. All models include only the dominant gravitational-wave mode. Investigations have been made into incorporating models with higher order modes into gravitational-wave searches~\citep{Capano:2013raa,Harry:2017weg}.

\subsection{Multi-detector Candidates and Significance}\label{sec:multi-det-candidates}

A ranking statistic is assigned to each potential candidate following the procedure in~\cite{Davies:2020}. The statistical significance of any given candidate is assessed by empirically estimating the rate of false alarms at the ranking statistic value associated with a candidate, and is typically reported as an inverse false alarm rate (IFAR). The distribution of false alarms is determined by the creation of numerous analyses which do not contain astrophysical candidates~\citep{Babak:2012zx,Usman:2015kfa}. This is achieved by analyzing the data set with time offsets between the detectors large enough to break the time-of-flight requirements for a true astrophysical signal. This procedure has been used successfully in many past analyses~\citep{Nitz:2019hdf,Nitz:2018imz,LIGOScientific:2018mvr,Venumadhav:2019tad,Colaboration:2011np,Abbott:2009qj,Abbott:2020niy}. Note, however, that this method is only applicable when multiple detectors are observing.

The IFAR of the search at the ranking statistic of a given candidate, however, does not answer the question of how likely a given candidate is to be astrophysical in origin, but rather the rate at which the search will produce candidates as statistically significant under the null hypothesis. For candidates which lie in part of the parameter space where a population model can be sufficiently described, as is the case for our focused BBH region, we can predict the rate of astrophysical sources and the distribution of true astrophysical sources which would be observed by our search for a given merger rate. The response of the search to a population of sources is directly measured by adding simulated gravitational-wave signals to the data. We model the full behavior of the search using a two-component mixture model of the expected astrophysical distribution and the empirically measured distribution of false alarms~\citep{Farr:2015}. A similar procedure has been used in past analyses of gravitational-wave data to assign the probability of astrophysical origin, or $\mathcal{P}_{\textrm{astro}}$~\citep{Nitz:2019hdf,Abbott:2020niy}. For multi-detector candidates which lie outside of the focused BBH region, in regions where the population of candidates is less certain or unknown, we choose not to assign a probability of astrophysical origin.

\subsection{Single-detector Candidates}

 In this catalog, we conduct a single-detector analysis
 of the focused BBH and BNS regions. We rely on the methods introduced in~\cite{Nitz:2020naa} to assess the probability of astrophysical origin of observed candidates. We assess the expected signal distribution in the same manner as for multi-detector candidates. However, due to the inability to empirically estimate the noise distribution for occurrences rarer than once per observing period, an extrapolation is needed;~\cite{Nitz:2020naa} introduces a purposefully conservative noise model for this purpose. Due to the mismatch in sensitive range between the LIGO and Virgo instruments (factor of 2-3x), we apply the single-detector analysis to time when a single LIGO observatory is operating, irrespective of Virgo's observing status. In order to limit the effects of possible astrophysical contamination, we assess the background using only data collected when both LIGO observatories were observing. This ensures that most strong astrophysical signals can be excised from the data using the multi-detector coincidence analysis first.

\section{Parameter Inference}

We infer the properties of BBH and BNS mergers by performing Bayesian analysis with the help of PyCBC Inference~\citep{Biwer:2018osg}. For BBHs, we use the latest version of the IMRPhenomXPHM waveform model~\citep{Pratten:2020ceb,lalsuite}, which includes sub-dominant harmonics and effects of precession on a quasi-circular BBH merger. In a recent study, this waveform model was used for doing parameter estimation on events from the first and second observing runs \citep{mateulucena2021adding}. We use the dynamical nested sampling algorithm \citep{Higson_2018,10.1214/06-BA127} implemented in the Dynesty software package \citep{speagle:2019} to sample over the parameter space, which includes chirp mass, mass ratio, spins (radial, polar, and azimuthal), distance, inclination angle, right ascension, declination,  coalescence phase, and the merger time. To help with sampler convergence we numerically marginalize over polarization.

For each of the events, we use uniform priors on source-frame component masses and merger time. We also assume a distance prior that is uniform in comoving volume; the luminosity distance ($D_L$) is related to the comoving volume assuming a flat $\Lambda$CDM cosmological model~\citep{Ade:2015xua}. An isotropic distribution of prior in the sky localization and binary orientation is assumed for each of the events. For the spins, we use uniform priors for the magnitude of the spin and isotropic for the orientation.

A low frequency cutoff ($f_{\textrm{low}}$) of 20 Hz is used for the evaluation of the likelihood function for all the detectors and for analyzing all events except for GW190727\_060333 ($f_{\textrm{low}}$ = 50 Hz for LIGO Livingston), GW190814\_211039 ($f_{\textrm{low}}$ = 30 Hz for LIGO Hanford) \citep{Abbott:2020niy}, and GW190725\_174728 ($f_{\textrm{low}} = 41\,$Hz for all detectors). In some instances, the raw data contains glitches as described in \citep{Abbott:2020niy}. Where available, we use the public glitch-subtracted data (for e.g. GW190413\_134308, GW190424\_180648, GW190425\_081805, GW190503\_185404, GW190513\_205428, GW190514\_065416, GW190701\_203306, and GW190924\_021846)~\citep{Vallisneri:2014vxa,Abbott:2019ebz}. We use 512 seconds of data around each candidate (with the exception of GW190725\_174728) to estimate the local power spectral density (PSD) using a version of Welch's method~\citep{findchirp}. The data is divided into overlapping segments (8s duration for BBH; 16s for BNS) which are Hann-windowed. The final PSD estimate is the result of taking the mean of the median average of the odd and even segments' power spectrum.

For GW190725\_174728, only 19 seconds of data were available in the Hanford detector prior to the event. In order to accommodate the longest signal possible from our prior volume in this time segment, it was necessary to use a lower frequency cutoff of $41\,$Hz. In addition, there was a glitch in the Hanford detector at $357$ seconds after GW190725\_174728. For this reason, we restricted the PSD estimation window for the Hanford detector to 360 seconds for this event.

For BNS mergers, we use the IMRPhenomD\_NRTidal waveform model \citep{PhysRevD.93.044007,PhysRevD.93.044006,PhysRevD.96.121501,PhysRevD.99.024029,lalsuite}, which includes tidal deformability parameters $\Lambda_1$ and $\Lambda_2$ of the two component masses. We use similar priors to that of the BBH analyses on component masses, comoving volume, merger time, and orientation. We use a heterodyne method ~\citep{Cornish:2010kf,Finstad:2020sok,zackay2018relative} to calculate the likelihood function. For the component spins, we assume spins aligned with the orbital angular moment with magnitude $\in [-0.05,0.05]$. We do not assume a common equation of state for the components; instead, we allow the tidal deformability of the components $\Lambda_{1,2}$ to vary independently of each other, using a prior uniform $\in [0, 5000]$ for both. A low frequency cutoff of 20 Hz is used to estimate the likelihood function.

Samplers based on nested sampling algorithms make use of ``live-points''. These are initially drawn from the prior volume, then slowly converge towards higher likelihood regions. In the process, the sampler estimates the Bayesian evidence ($Z$), which is defined as the integral of the likelihood times the prior volume. We let the sampler run until the estimated remaining log-evidence is equal to a predefined value, which we set to 0.1. Where possible, we compare our results to posteriors from previous catalogs \citep{Nitz:2019hdf,LIGOScientific:2018mvr, Abbott:2020niy}, considering the sampler converged if the posteriors agree up to differences expected due to the use of updated waveform models. For most events we find it sufficient to use 4000 live points. However, for a few events, it was necessary to increase the number of live points to obtain converged posteriors. For these events we increased the number of live points by 4000 until we found the posteriors to be the same between two resolutions. This resulted in between 8000 and 20000 live points used for these events.

\section{Observational Results}

From the combined analysis of the 2015-2019 public LIGO and Virgo data, we find 55 BBH mergers and 2 BNS mergers. The list of gravitational wave mergers is given in Table~\ref{table:search}. For the majority of BBHs we can assess the probability of astrophysical origin. Our catalog includes candidates where $\mathcal{P}_{\textrm{astro}} > 0.5$ or $\mathrm{IFAR} > 100$ years. These thresholds are consistent with prior community conventions~\cite{LIGOScientific:2016dsl,LIGOScientific:2018mvr, Nitz:2018imz,Nitz:2019hdf,Venumadhav:2019lyq}. The marginalized parameter estimates for source-frame component masses, chirp mass, mass ratio, effective spin, luminosity distance, redshift, final mass, and final spin obtained from the posterior distributions are listed in table \ref{table:peresults}. 

\begin{figure*}[t]
  \centering
    \includegraphics[width=1.05\columnwidth]{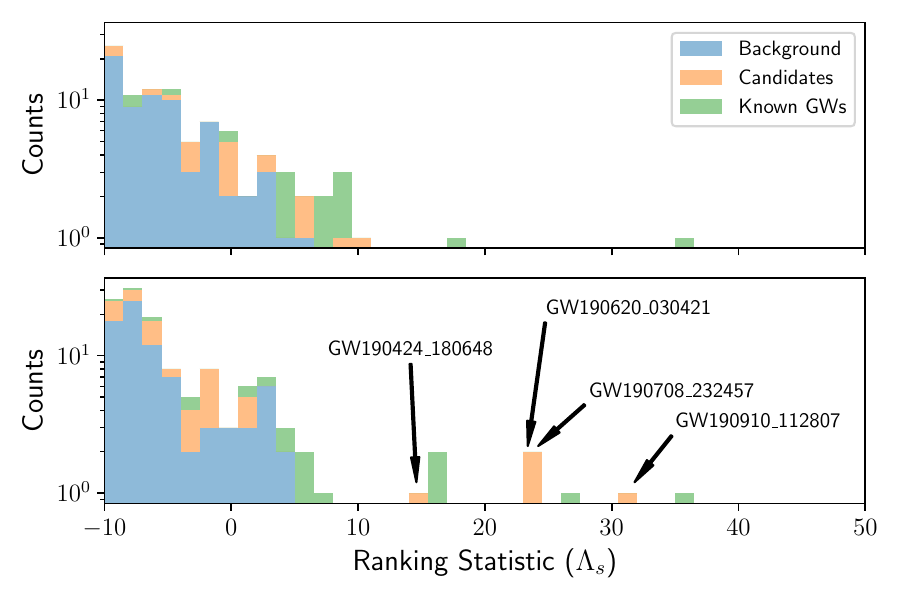}
        \includegraphics[width=1.05\columnwidth]{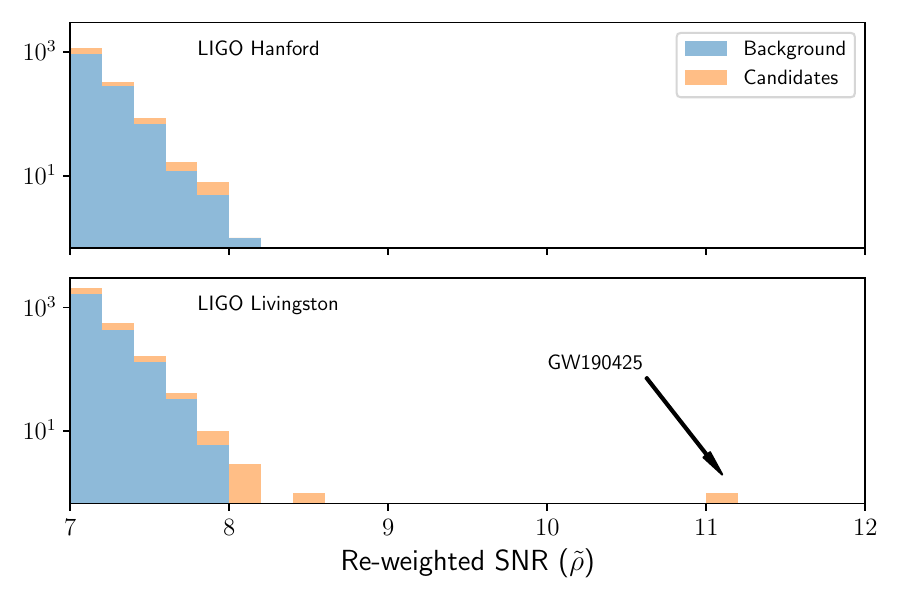}
\caption{The stacked distributions of single-detector triggered candidates observed when a single LIGO observatory was operating (green), our selected background (blue), and for comparison the distribution of gravitational-wave mergers observed by the multi-detector analysis (orange) as a function of the ranking statistic. To estimate the significance of the candidates, the method of~\cite{Nitz:2020naa} is used to extrapolate the background distribution, which allows us to estimate the probability of astrophysical origin. Shown are the results of the BBH analysis (left), which uses the statistic $\lambda_s$~\citep{Nitz:2020naa}, and BNS analysis (right), which uses a re-weighted SNR statistic~\citep{Babak:2012zx,Nitz:2019hdf}, for the LIGO Hanford (top) and LIGO Livingston (bottom) data during O3a.}
\label{fig:single}
\end{figure*}

Several candidates were independently detected by Virgo, with the Virgo observatory being decisive in the case of two of them. As the gap in sensitivity between the LIGO and Virgo instruments narrows, we expect this to become more commonplace. We identify four candidates in our single-detector analysis of BNS and BBH mergers in LIGO Hanford and LIGO Livingston data as shown in Fig.~\ref{fig:single}. These are consistent with the previously-reported single-detector analysis of~\cite{Abbott:2020niy}. 

We find four previously unreported BBH mergers; three, GW190725\_174728, GW190916\_200658, and GW190926\_050336 are near-threshold observations with relatively low SNR. The fourth, GW190925\_232845, has SNR $\sim 10$ and is found at a false alarm rate  $<1 / 100~ \textrm{years}$. We find GW190925\_232845 has component masses $20.2^{+3.9}_{-2.5}\msun$ and $15.6^{+2.1}_{-2.6}\msun$. While not reported as a new BBH merger detection, this time was noted as part of a recent search for lensed images~\citep{McIsaac:2019use, Abbott:2021iab}. The remainder of the multi-detector observed mergers are broadly consistent with previous searches~\citep{Venumadhav:2019lyq,Abbott:2020niy,LIGOScientific:2018mvr}. Two marginal observations reported in~\cite{Abbott:2020niy}, 190426\_152155 and 190909\_114149, are not assigned high significance in our analysis, but notably, our updated catalog now includes two candidates which were originally reported in~\cite{Venumadhav:2019lyq} from O2, GW170202\_135657 and GW170403\_230611.

\setlength{\LTcapwidth}{\textwidth}
\begin{longtable*}{cccccccccc}
  \caption{Gravitational-wave observations from the full search of O1-O3a data with $\mathcal{P}_{\textrm{astro}} > 0.5$ or IFAR $>$ 100 years. Candidates are sorted by observation time. For each candidate, we show the detectors that were observing at the time, the subset which triggered on the event within our analysis, and the SNR ($\rho$) reported by the search for each detector. Due to thresholds on the SNR and the ability for the search to select a preferred candidate from many at a given time, there may be no detector SNR associated with a candidate, even if it is observing at the time. For multi-detector candidates, we show the false alarm rate of the entire search at the threshold of its ranking statistic value. For BBHs found by our focused BBH search, we give estimates of the probability of astrophysical origin  $\mathcal{P}_\textrm{astro}$. We also show our estimates for single-detector candidates, which we note will necessarily be more uncertain, due to the need to extrapolate the background model. GW190425 is assessed using the same conservative extrapolation of the background as for BBH candidates, however, we expect that the noise distribution may be more well-behaved than assumed here for such a long duration signal. Candidates reported here for the first time are in bold.}
\label{table:search}\\
 & Event & GPS Time & Observing & Triggered & $\mathcal{P}_\textrm{astro}$ & IFAR [yr] & $\rho_H$ &  $\rho_L$  & $\rho_V$  \\\hline
    \endfirsthead
  \caption*{(Continued) Gravitational-wave observations from the full search of O1-O3a data with $\mathcal{P}_{\textrm{astro}} > 0.5$ or IFAR $>$ 100 years. Candidates are sorted by observation time. For each candidate, we show the detectors that were observing at the time, the subset which triggered on the event within our analysis, and the SNR ($\rho$) reported by the search for each detector. Due to thresholds on the SNR and the ability for the search to select a preferred candidate from many at a given time, there may be no detector SNR associated with a candidate, even if it is observing at the time. For multi-detector candidates, we show the false alarm rate of the entire search at the threshold of its ranking statistic value. For BBHs found by our focused BBH search, we give estimates of the probability of astrophysical origin  $\mathcal{P}_\textrm{astro}$. We also show our estimates for single-detector candidates, which we note will necessarily be more uncertain, due to the need to extrapolate the background model. GW190425 is assessed using the same conservative extrapolation of the background as for BBH candidates, however, we expect that the noise distribution may be more well-behaved than assumed here for such a long duration signal. Candidates reported here for the first time are in bold.}
      \label{table:search}\\
 & Event & GPS Time & Observing & Triggered & $\mathcal{P}_\textrm{astro}$ & IFAR [yr] & $\rho_H$ &  $\rho_L$  & $\rho_V$  \\\hline
    \endhead 
    \hline
    \endfoot
 1 & GW150914\_095045 & 1126259462.43 & HL & HL & 1.00 & $>100$ & 19.9 & 13.0 & - \\
\rowcolor{gray!20} 2 & GW151012\_095443 & 1128678900.45 & HL & HL & 1.00 & $>100$ & 6.9 & 6.6 & - \\
\rowcolor{white} 3 & GW151226\_033853 & 1135136350.65 & HL & HL & 1.00 & $>100$ & 10.5 & 7.4 & - \\
\rowcolor{gray!20} 4 & GW170104\_101158 & 1167559936.60 & HL & HL & 1.00 & $>100$ & 8.9 & 9.6 & - \\
\rowcolor{white} 5 & GW170121\_212536 & 1169069154.58 & HL & HL & 1.00 & 16 & 5.2 & 8.9 & - \\
\rowcolor{gray!20} 6 & GW170202\_135657 & 1170079035.73 & HL & HL & 0.81 & 0.50 & 5.4 & 6.2 & - \\
\rowcolor{white} 7 & GW170304\_163753 & 1172680691.37 & HL & HL & 0.70 & 0.25 & 4.6 & 7.0 & - \\
\rowcolor{gray!20} 8 & GW170403\_230611 & 1175295989.23 & HL & HL & 0.71 & 0.25 & 5.2 & 5.5 & - \\
\rowcolor{white} 9 & GW170608\_020116 & 1180922494.49 & HL & HL & 1.00 & $>100$ & 12.4 & 9.0 & - \\
\rowcolor{gray!20} 10 & GW170727\_010430 & 1185152688.03 & HL & HL & 1.00 & 71 & 4.7 & 7.5 & - \\
\rowcolor{white} 11 & GW170729\_185629 & 1185389807.32 & HL & HL & 0.99 & 28 & 7.5 & 6.9 & - \\
\rowcolor{gray!20} 12 & GW170809\_082821 & 1186302519.75 & HLV & HL & 1.00 & $>100$ & 6.7 & 10.7 & - \\
\rowcolor{white} 13 & GW170814\_103043 & 1186741861.53 & HLV & HL & 1.00 & $>100$ & 9.2 & 13.7 & - \\
\rowcolor{gray!20} 14 & GW170817\_124104 & 1187008882.45 & HLV & HL & - & $>100$ & 18.3 & 25.5 & - \\
\rowcolor{white} 15 & GW170818\_022509 & 1187058327.08 & HLV & HL & 1.00 & 5.26 & 4.5 & 9.6 & - \\
\rowcolor{gray!20} 16 & GW170823\_131358 & 1187529256.52 & HL & HL & 1.00 & $>100$ & 6.6 & 9.1 & - \\
\rowcolor{white} 17 & GW190408\_181802 & 1238782700.28 & HLV & HL & 1.00 & $>100$ & 9.2 & 10.3 & - \\
\rowcolor{gray!20} 18 & GW190412\_053044 & 1239082262.17 & HLV & HL & 1.00 & $>100$ & 8.2 & 14.9 & - \\
\rowcolor{white} 19 & GW190413\_052954 & 1239168612.50 & HLV & HL & 0.99 & 1.45 & 5.2 & 6.7 & - \\
\rowcolor{gray!20} 20 & GW190413\_134308 & 1239198206.74 & HLV & HL & 0.99 & 6.39 & 5.4 & 7.8 & - \\
\rowcolor{white} 21 & GW190421\_213856 & 1239917954.25 & HL & HL & 1.00 & $>100$ & 7.9 & 6.3 & - \\
\rowcolor{gray!20} 22 & GW190424\_180648 & 1240164426.14 & L & L & 0.81 & - & - & 9.9 & - \\
\rowcolor{white} 23 & GW190425\_081805 & 1240215503.02 & LV & L & 0.50 & - & - & 11.9 & - \\
\rowcolor{gray!20} 24 & GW190503\_185404 & 1240944862.29 & HLV & HL & 1.00 & $>100$ & 9.1 & 7.6 & - \\
\rowcolor{white} 25 & GW190512\_180714 & 1241719652.42 & HLV & HL & 1.00 & $>100$ & 5.9 & 10.8 & - \\
\rowcolor{gray!20} 26 & GW190513\_205428 & 1241816086.74 & HLV & HLV & 1.00 & $>100$ & 8.8 & 7.7 & 4.0 \\
\rowcolor{white} 27 & GW190514\_065416 & 1241852074.85 & HL & HL & 0.85 & 0.19 & 6.1 & 5.3 & - \\
\rowcolor{gray!20} 28 & GW190517\_055101 & 1242107479.83 & HLV & HL & 1.00 & 66 & 6.8 & 7.9 & - \\
\rowcolor{white} 29 & GW190519\_153544 & 1242315362.38 & HLV & HL & 1.00 & $>100$ & 7.8 & 9.3 & - \\
\rowcolor{gray!20} 30 & GW190521\_030229 & 1242442967.44 & HLV & HL & 1.00 & $>100$ & 8.4 & 12.0 & - \\
\rowcolor{white} 31 & GW190521\_074359 & 1242459857.47 & HL & HL & 1.00 & $>100$ & 12.1 & 21.0 & - \\
\rowcolor{gray!20} 32 & GW190527\_092055 & 1242984073.79 & HL & HL & 0.93 & 0.37 & 5.0 & 7.0 & - \\
\rowcolor{white} 33 & GW190602\_175927 & 1243533585.10 & HLV & HL & 1.00 & $>100$ & 6.2 & 10.8 & - \\
\rowcolor{gray!20} 34 & GW190620\_030421 & 1245035079.31 & LV & L & 0.85 & - & - & 11.2 & - \\
\rowcolor{white} 35 & GW190630\_185205 & 1245955943.18 & LV & LV & 1.00 & 0.18 & - & 14.7 & 4.0 \\
\rowcolor{gray!20} 36 & GW190701\_203306 & 1246048404.58 & HLV & HLV & 1.00 & 0.13 & 6.0 & 8.9 & 5.7 \\
\rowcolor{white} 37 & GW190706\_222641 & 1246487219.33 & HLV & HL & 1.00 & $>100$ & 9.4 & 8.6 & - \\
\rowcolor{gray!20} 38 & GW190707\_093326 & 1246527224.17 & HL & HL & 1.00 & $>100$ & 7.9 & 9.6 & - \\
\rowcolor{white} 39 & GW190708\_232457 & 1246663515.38 & LV & L & 0.85 & - & - & 12.6 & - \\
\rowcolor{gray!20} 40 & GW190719\_215514 & 1247608532.92 & HL & HL & 0.89 & 0.25 & 5.6 & 5.7 & - \\
\rowcolor{white} 41 & GW190720\_000836 & 1247616534.71 & HLV & HL & 1.00 & $>100$ & 6.8 & 7.7 & - \\
\rowcolor{gray!20} 42 & \textbf{GW190725\_174728} & 1248112066.46 & HLV & HL & 0.91 & 0.41 & 5.4 & 7.3 & - \\
\rowcolor{white} 43 & GW190727\_060333 & 1248242631.98 & HLV & HL & 1.00 & $>100$ & 7.9 & 8.1 & - \\
\rowcolor{gray!20} 44 & GW190728\_064510 & 1248331528.53 & HLV & HL & 1.00 & $>100$ & 7.5 & 10.6 & - \\
\rowcolor{white} 45 & GW190731\_140936 & 1248617394.64 & HL & HL & 0.93 & 0.43 & 5.2 & 6.0 & - \\
\rowcolor{gray!20} 46 & GW190803\_022701 & 1248834439.88 & HLV & HL & 0.99 & 2.40 & 5.6 & 6.7 & - \\
\rowcolor{white} 47 & GW190814\_211039 & 1249852257.01 & HLV & HL & - & $>100$ & 11.0 & 21.1 & - \\
\rowcolor{gray!20} 48 & GW190828\_063405 & 1251009263.76 & HLV & HL & 1.00 & $>100$ & 10.3 & 11.2 & - \\
\rowcolor{white} 49 & GW190828\_065509 & 1251010527.89 & HLV & HL & 1.00 & $>100$ & 7.3 & 7.4 & - \\
\rowcolor{gray!20} 50 & GW190910\_112807 & 1252150105.32 & LV & L & 0.87 & - & - & 13.4 & - \\
\rowcolor{white} 51 & GW190915\_235702 & 1252627040.70 & HLV & HL & 1.00 & $>100$ & 9.0 & 8.6 & - \\
\rowcolor{gray!20} 52 & \textbf{GW190916\_200658} & 1252699636.90 & HLV & HL & 0.88 & 0.22 & 4.9 & 5.9 & - \\
\rowcolor{white} 53 & GW190924\_021846 & 1253326744.84 & HLV & HL & 1.00 & $>100$ & 6.7 & 10.8 & - \\
\rowcolor{gray!20} 54 & \textbf{GW190925\_232845} & 1253489343.12 & HV & HV & 1.00 & $>100$ & 8.2 & - & 5.4 \\
\rowcolor{white} 55 & \textbf{GW190926\_050336} & 1253509434.07 & HLV & HL & 0.88 & 0.27 & 5.4 & 5.6 & - \\
\rowcolor{gray!20} 56 & GW190929\_012149 & 1253755327.50 & HLV & HL & 0.98 & 3.08 & 5.8 & 7.4 & - \\
\rowcolor{white} 57 & GW190930\_133541 & 1253885759.24 & HL & HL & 1.00 & $>100$ & 6.7 & 7.4 & - \\
     %\hline
\end{longtable*}

\begin{longtable*}{ccccccccccccc}
  \caption{The selection of sub-threshold candidates with $\mathcal{P}_{\textrm{astro}} > 0.2$ or IFAR $ > 0.5$ from the full search of O1-O3a data. Candidates are sorted by the observation time. The complete set of sub-threshold candidate is available in the data release and includes a selection of full parameter estimates. Here we show the detector-frame (redshifted) parameters of the template which triggered on the candidate, along with the reported SNRs ($\rho$) from each detector.}
      \label{table:sub}\\
 & Event & GPS Time & Observing & Triggered & $\mathcal{P}_\textrm{astro}$ & IFAR & $\rho_H$ &  $\rho_L$  & $\rho_V$ & $m_{1,\textrm{det}}/\msun$ & $m_{2,\textrm{det}}/\msun$ & $\chi_\mathrm{eff}$\\\hline
 \endfirsthead
   \caption*{(Continued) The selection of sub-threshold candidates with $\mathcal{P}_{\textrm{astro}} > 0.2$ or IFAR $ > 0.5$ from the full search of O1-O3a data. Candidates are sorted by the observation time. The complete set of sub-threshold candidate is available in the data release and includes a selection of full parameter estimates. Here we show the detector-frame (redshifted) parameters of the template which triggered on the candidate, along with the reported SNRs ($\rho$) from each detector.}
    \endhead 
    \hline
    \endfoot
 1 & 151011\_192749 & 1128626886.60 & HL & HL & 0.21 & 0.02 & 4.7 & 6.8 & - & 33.5 & 65.6 & 0.1\\
\rowcolor{gray!20} 2 & 151205\_195525 & 1133380542.41 & HL & HL & 0.25 & 0.03 & 5.9 & 4.8 & - & 81.6 & 77.7 & 0.1\\
\rowcolor{white} 3 & 170425\_055334 & 1177134832.19 & HL & HL & 0.41 & 0.07 & 5.3 & 5.8 & - & 46.1 & 65.0 & 0.1\\
\rowcolor{gray!20} 4 & 170704\_202003 & 1183234821.62 & HL & HL & 0.34 & 0.05 & 5.1 & 6.5 & - & 10.0 & 13.2 & -0.0\\
\rowcolor{white} 5 & 170722\_065503 & 1184741721.32 & HL & HL & - & 0.89 & 5.0 & 7.3 & - & 1.7 & 1.3 & -0.0\\
\rowcolor{gray!20} 6 & 190404\_142514 & 1238423132.99 & HL & HL & 0.44 & 0.02 & 5.1 & 5.9 & - & 22.5 & 24.5 & 0.1\\
\rowcolor{white} 7 & 190426\_053949 & 1240292407.21 & HLV & HL & 0.32 & 0.01 & 5.2 & 6.1 & - & 20.7 & 20.0 & 0.2\\
\rowcolor{gray!20} 8 & 190427\_180650 & 1240423628.68 & HLV & HL & 0.41 & 0.02 & 5.8 & 6.8 & - & 13.0 & 7.9 & -0.0\\
\rowcolor{white} 9 & 190509\_004120 & 1241397698.79 & HLV & HL & 0.31 & 0.01 & 4.7 & 6.2 & - & 30.1 & 28.2 & -0.0\\
\rowcolor{gray!20} 10 & 190524\_134109 & 1242740487.36 & HLV & HL & 0.21 & 0.01 & 4.3 & 6.0 & - & 123.3 & 77.2 & 0.2\\
\rowcolor{white} 11 & 190530\_030659 & 1243220837.97 & HLV & HL & 0.31 & 0.01 & 5.2 & 5.8 & - & 26.3 & 45.4 & 0.2\\
\rowcolor{gray!20} 12 & 190630\_135302 & 1245938000.49 & HL & HL & 0.23 & 0.01 & 5.1 & 5.8 & - & 32.6 & 19.2 & 0.0\\
\rowcolor{white} 13 & 190704\_104834 & 1246272532.92 & HLV & HL & 0.26 & 0.01 & 7.0 & 5.5 & - & 5.0 & 5.4 & 0.1\\
\rowcolor{gray!20} 14 & 190707\_071722 & 1246519060.10 & HLV & HL & 0.21 & 0.01 & 6.0 & 5.7 & - & 10.7 & 14.1 & 0.0\\
\rowcolor{white} 15 & 190805\_105432 & 1249037690.78 & HL & HL & 0.41 & 0.02 & 4.8 & 6.5 & - & 9.4 & 18.3 & -0.1\\
\rowcolor{gray!20} 16 & 190808\_230535 & 1249340753.59 & HLV & HL & 0.31 & 0.01 & 5.0 & 6.5 & - & 13.6 & 13.6 & 0.2\\
\rowcolor{white} 17 & 190821\_050019 & 1250398837.88 & HLV & HL & 0.23 & 0.01 & 5.2 & 5.6 & - & 26.8 & 17.0 & -0.1\\
    %\hline
\end{longtable*}

\setlength{\tabcolsep}{1mm}
\begin{longtable*}{cccccccccccccc}
  \caption{Bayesian parameter estimation for the 57 detections in the entire O1-O3a data. We report the median value and $90\%$ credible interval for the source-frame component mass $m_1$ and $m_2$, chirp mass $\mathcal{M}$, mass ratio $q$, effective spin $\chi_\mathrm{eff}$, luminosity distance $D_\mathrm{L}$, redshift $z$, and remnant mass and spin $M_f$ and $\chi_f$, respectively. The signal-to-noise ratio (SNR) is computed from the maximum likelihood with polarization angle being numerically marginalized for BBH events and with phase analytically marginalized for BNS events. Candidates reported here for the first time are in bold.}\\
  &Event 
& $ {m_1}/M_\odot $ 
& ${m_2}/M_\odot $ 
& $\mathcal{M}/M_\odot $ 
& $q$ 
& $\chi_\mathrm{eff}$ 
& ${D_\mathrm{L}}/\mathrm{Mpc} $ 
& $z$ 
& ${M_\mathrm{f}}/M_\odot$ 
& $\chi_f$ & SNR  \\\hline
  \endfirsthead
   \caption*{(Continued) Bayesian parameter estimation for the 57 detections in the entire O1-O3a data. We report the median value and $90\%$ credible interval for the source-frame component mass $m_1$ and $m_2$, chirp mass $\mathcal{M}$, mass ratio $q$, effective spin $\chi_\mathrm{eff}$, luminosity distance $D_\mathrm{L}$, redshift $z$, and remnant mass and spin $M_f$ and $\chi_f$, respectively. The signal-to-noise ratio (SNR) is computed from the maximum likelihood with polarization angle being numerically marginalized for BBH events and with phase analytically marginalized for BNS events. Candidates reported here for the first time are in bold.}
      \label{table:peresults}\\
&Event 
& $ {m_1}/M_\odot $ 
& ${m_2}/M_\odot $ 
& $\mathcal{M}/M_\odot $ 
& $q$ 
& $\chi_\mathrm{eff}$ 
& ${D_\mathrm{L}}/\mathrm{Mpc} $ 
& $z$ 
& ${M_\mathrm{f}}/M_\odot$ 
& $\chi_f$ & SNR  \\\hline
    \endhead 
    \hline
    \endfoot
 1& $\mathrm{GW150914\_095045}$ & $34.7^{+4.7}_{-2.8}$ & $29.8^{+2.8}_{-4.4}$ & $27.9^{+1.4}_{-1.3}$ & $1.2^{+0.4}_{-0.1}$ & $-0.03^{+0.11}_{-0.13}$ & $534^{+123}_{-157}$ & $0.11^{+0.02}_{-0.03}$ & $61.5^{+2.9}_{-2.7}$ & $0.67^{+0.03}_{-0.05}$ & 23.8 \\ 
\rowcolor{Gray}2& $\mathrm{GW151012\_095443}$ & $27.0^{+10.5}_{-7.9}$ & $11.8^{+4.3}_{-3.1}$ & $15.2^{+1.1}_{-0.9}$ & $2.3^{+2.0}_{-1.1}$ & $0.05^{+0.22}_{-0.15}$ & $927^{+456}_{-329}$ & $0.18^{+0.08}_{-0.06}$ & $37.4^{+7.7}_{-4.8}$ & $0.62^{+0.08}_{-0.08}$ & 9.8 \\ 
3& $\mathrm{GW151226\_033853}$ & $14.1^{+10.3}_{-3.4}$ & $7.4^{+2.2}_{-2.6}$ & $8.8^{+0.2}_{-0.2}$ & $1.9^{+3.2}_{-0.8}$ & $0.22^{+0.21}_{-0.08}$ & $503^{+133}_{-156}$ & $0.11^{+0.03}_{-0.03}$ & $20.5^{+7.9}_{-1.5}$ & $0.72^{+0.03}_{-0.03}$ & 13.2 \\ 
\rowcolor{Gray}4& $\mathrm{GW170104\_101158}$ & $29.1^{+6.3}_{-4.4}$ & $20.5^{+4.2}_{-4.3}$ & $21.0^{+1.9}_{-1.5}$ & $1.4^{+0.7}_{-0.4}$ & $-0.06^{+0.15}_{-0.18}$ & $1077^{+384}_{-458}$ & $0.21^{+0.07}_{-0.08}$ & $47.5^{+4.2}_{-3.3}$ & $0.65^{+0.06}_{-0.09}$ & 13.7 \\ 
5& $\mathrm{GW170121\_212536}$ & $33.0^{+8.7}_{-5.6}$ & $25.1^{+5.6}_{-6.4}$ & $24.9^{+3.7}_{-3.2}$ & $1.3^{+0.8}_{-0.3}$ & $-0.19^{+0.23}_{-0.29}$ & $1164^{+981}_{-673}$ & $0.23^{+0.16}_{-0.12}$ & $55.9^{+7.9}_{-7.0}$ & $0.61^{+0.08}_{-0.13}$ & 10.9 \\ 
\rowcolor{Gray}6& $\mathrm{GW170202\_135657}$ & $30.9^{+12.6}_{-8.2}$ & $13.7^{+6.3}_{-3.8}$ & $17.5^{+3.1}_{-1.5}$ & $2.3^{+2.1}_{-1.1}$ & $-0.08^{+0.29}_{-0.31}$ & $1422^{+735}_{-608}$ & $0.27^{+0.12}_{-0.11}$ & $43.6^{+9.7}_{-5.6}$ & $0.54^{+0.17}_{-0.15}$ & 8.5 \\ 
7& $\mathrm{GW170304\_163753}$ & $44.5^{+14.7}_{-9.0}$ & $31.7^{+9.3}_{-11.2}$ & $32.1^{+6.8}_{-5.5}$ & $1.4^{+1.3}_{-0.3}$ & $0.1^{+0.27}_{-0.26}$ & $2354^{+1584}_{-1293}$ & $0.42^{+0.22}_{-0.21}$ & $72.4^{+15.0}_{-10.8}$ & $0.7^{+0.09}_{-0.13}$ & 8.7 \\ 
\rowcolor{Gray}8& $\mathrm{GW170403\_230611}$ & $49.1^{+16.4}_{-10.5}$ & $35.8^{+11.4}_{-12.4}$ & $35.9^{+8.7}_{-6.7}$ & $1.3^{+1.1}_{-0.3}$ & $-0.21^{+0.33}_{-0.35}$ & $2967^{+2204}_{-1540}$ & $0.51^{+0.29}_{-0.24}$ & $81.4^{+19.0}_{-14.0}$ & $0.6^{+0.11}_{-0.17}$ & 7.7 \\ 
9& $\mathrm{GW170608\_020116}$ & $10.6^{+3.1}_{-1.3}$ & $8.0^{+1.1}_{-1.7}$ & $8.0^{+0.2}_{-0.2}$ & $1.3^{+0.9}_{-0.3}$ & $0.07^{+0.11}_{-0.06}$ & $321^{+129}_{-110}$ & $0.07^{+0.03}_{-0.02}$ & $17.8^{+1.5}_{-0.6}$ & $0.69^{+0.02}_{-0.03}$ & 15.2 \\ 
\rowcolor{Gray}10& $\mathrm{GW170727\_010430}$ & $41.3^{+12.1}_{-7.6}$ & $30.8^{+8.0}_{-8.9}$ & $30.7^{+5.5}_{-4.6}$ & $1.3^{+0.9}_{-0.3}$ & $-0.04^{+0.23}_{-0.3}$ & $2153^{+1529}_{-1050}$ & $0.39^{+0.22}_{-0.17}$ & $69.1^{+11.9}_{-9.7}$ & $0.66^{+0.08}_{-0.14}$ & 8.8 \\ 
11& $\mathrm{GW170729\_185629}$ & $53.8^{+12.1}_{-11.8}$ & $31.6^{+13.0}_{-10.4}$ & $35.1^{+7.9}_{-5.8}$ & $1.7^{+1.1}_{-0.6}$ & $0.28^{+0.22}_{-0.29}$ & $2236^{+1590}_{-1215}$ & $0.4^{+0.23}_{-0.2}$ & $80.9^{+15.0}_{-11.0}$ & $0.76^{+0.08}_{-0.18}$ & 10.8 \\ 
\rowcolor{Gray}12& $\mathrm{GW170809\_082821}$ & $33.9^{+8.2}_{-5.0}$ & $24.5^{+4.7}_{-5.5}$ & $24.9^{+2.1}_{-1.6}$ & $1.4^{+0.8}_{-0.3}$ & $0.08^{+0.16}_{-0.16}$ & $1069^{+293}_{-364}$ & $0.21^{+0.05}_{-0.07}$ & $55.8^{+4.8}_{-3.5}$ & $0.69^{+0.06}_{-0.08}$ & 12.4 \\ 
13& $\mathrm{GW170814\_103043}$ & $30.9^{+5.4}_{-3.2}$ & $24.9^{+2.9}_{-4.0}$ & $24.0^{+1.3}_{-1.0}$ & $1.2^{+0.5}_{-0.2}$ & $0.07^{+0.12}_{-0.12}$ & $593^{+149}_{-207}$ & $0.12^{+0.03}_{-0.04}$ & $53.2^{+3.0}_{-2.3}$ & $0.7^{+0.04}_{-0.04}$ & 17.4 \\ 
\rowcolor{Gray}14& $\mathrm{GW170817\_124104}$ & $1.4^{+0.1}_{-0.1}$ & $1.3^{+0.1}_{-0.1}$ & $1.186^{+0.003}_{-0.001}$ & $1.1^{+0.2}_{-0.1}$ & $-0.0^{+0.01}_{-0.01}$ & $43^{+5}_{-10}$ & $0.01^{+0.0}_{-0.0}$ &  - &  - & 32.7 \\ 
15& $\mathrm{GW170818\_022509}$ & $35.2^{+7.1}_{-4.6}$ & $27.0^{+4.4}_{-5.3}$ & $26.7^{+2.2}_{-1.9}$ & $1.3^{+0.6}_{-0.3}$ & $-0.07^{+0.19}_{-0.23}$ & $1073^{+434}_{-411}$ & $0.21^{+0.07}_{-0.07}$ & $59.6^{+4.7}_{-4.0}$ & $0.65^{+0.07}_{-0.09}$ & 11.8 \\ 
\rowcolor{Gray}16& $\mathrm{GW170823\_131358}$ & $38.1^{+9.7}_{-6.0}$ & $28.6^{+6.4}_{-7.5}$ & $28.4^{+4.2}_{-3.1}$ & $1.3^{+0.8}_{-0.3}$ & $0.05^{+0.2}_{-0.23}$ & $1965^{+810}_{-873}$ & $0.36^{+0.12}_{-0.14}$ & $63.5^{+9.0}_{-6.3}$ & $0.69^{+0.07}_{-0.11}$ & 11.4 \\ 
17& $\mathrm{GW190408\_181802}$ & $24.6^{+5.2}_{-3.4}$ & $18.4^{+3.4}_{-3.7}$ & $18.3^{+1.8}_{-1.2}$ & $1.3^{+0.6}_{-0.3}$ & $-0.04^{+0.13}_{-0.16}$ & $1585^{+447}_{-625}$ & $0.3^{+0.07}_{-0.11}$ & $41.1^{+3.9}_{-2.7}$ & $0.66^{+0.05}_{-0.07}$ & 14.0 \\ 
\rowcolor{Gray}18& $\mathrm{GW190412\_053044}$ & $30.4^{+5.8}_{-4.2}$ & $8.2^{+1.2}_{-1.1}$ & $13.2^{+0.5}_{-0.3}$ & $3.7^{+1.4}_{-0.9}$ & $0.25^{+0.12}_{-0.1}$ & $757^{+156}_{-200}$ & $0.15^{+0.03}_{-0.04}$ & $37.5^{+4.9}_{-3.2}$ & $0.65^{+0.04}_{-0.03}$ & 19.1 \\ 
19& $\mathrm{GW190413\_052954}$ & $34.7^{+11.2}_{-6.7}$ & $25.1^{+6.7}_{-7.3}$ & $25.4^{+4.8}_{-3.7}$ & $1.4^{+1.0}_{-0.3}$ & $-0.02^{+0.27}_{-0.33}$ & $3193^{+1784}_{-1366}$ & $0.54^{+0.24}_{-0.2}$ & $57.3^{+10.7}_{-7.9}$ & $0.66^{+0.09}_{-0.15}$ & 9.0 \\ 
\rowcolor{Gray}20& $\mathrm{GW190413\_134308}$ & $51.6^{+16.9}_{-12.7}$ & $31.2^{+11.4}_{-11.9}$ & $34.1^{+7.5}_{-6.4}$ & $1.6^{+1.6}_{-0.6}$ & $-0.01^{+0.27}_{-0.35}$ & $3835^{+2665}_{-1899}$ & $0.63^{+0.34}_{-0.27}$ & $79.4^{+16.0}_{-12.9}$ & $0.64^{+0.11}_{-0.25}$ & 9.9 \\ 
21& $\mathrm{GW190421\_213856}$ & $42.2^{+10.8}_{-7.8}$ & $31.4^{+8.5}_{-10.6}$ & $31.1^{+6.0}_{-5.1}$ & $1.3^{+1.1}_{-0.3}$ & $-0.06^{+0.23}_{-0.3}$ & $2679^{+1605}_{-1251}$ & $0.47^{+0.22}_{-0.19}$ & $70.1^{+12.3}_{-9.8}$ & $0.65^{+0.08}_{-0.15}$ & 9.9 \\ 
\rowcolor{Gray}22& $\mathrm{GW190424\_180648}$ & $40.2^{+10.9}_{-7.1}$ & $30.7^{+7.3}_{-8.4}$ & $30.3^{+5.1}_{-4.4}$ & $1.3^{+0.8}_{-0.3}$ & $0.09^{+0.22}_{-0.27}$ & $2134^{+1466}_{-1115}$ & $0.38^{+0.21}_{-0.18}$ & $67.6^{+10.9}_{-9.3}$ & $0.7^{+0.07}_{-0.12}$ & 10.3 \\ 
23& $\mathrm{GW190425\_081805}$ & $1.8^{+0.2}_{-0.1}$ & $1.5^{+0.1}_{-0.1}$ & $1.432^{+0.013}_{-0.013}$ & $1.2^{+0.3}_{-0.2}$ & $0.02^{+0.02}_{-0.02}$ & $174^{+46}_{-43}$ & $0.04^{+0.01}_{-0.01}$ &  - &  - & 12.4 \\ 
\rowcolor{Gray}24& $\mathrm{GW190503\_185404}$ & $42.5^{+10.9}_{-8.5}$ & $27.5^{+8.1}_{-8.7}$ & $29.2^{+4.8}_{-3.9}$ & $1.5^{+1.2}_{-0.5}$ & $-0.04^{+0.22}_{-0.29}$ & $1478^{+608}_{-619}$ & $0.28^{+0.1}_{-0.11}$ & $66.9^{+10.0}_{-7.3}$ & $0.64^{+0.09}_{-0.18}$ & 12.2 \\ 
25& $\mathrm{GW190512\_180714}$ & $23.2^{+6.1}_{-5.8}$ & $12.5^{+3.6}_{-2.7}$ & $14.6^{+1.4}_{-0.9}$ & $1.9^{+1.1}_{-0.8}$ & $0.04^{+0.14}_{-0.15}$ & $1499^{+481}_{-618}$ & $0.28^{+0.08}_{-0.11}$ & $34.4^{+4.3}_{-3.5}$ & $0.65^{+0.06}_{-0.07}$ & 12.0 \\ 
\rowcolor{Gray}26& $\mathrm{GW190513\_205428}$ & $35.2^{+10.7}_{-9.3}$ & $18.1^{+7.4}_{-5.0}$ & $21.5^{+3.4}_{-2.2}$ & $1.9^{+1.4}_{-0.9}$ & $0.14^{+0.23}_{-0.21}$ & $2195^{+833}_{-815}$ & $0.39^{+0.12}_{-0.13}$ & $51.4^{+8.1}_{-6.4}$ & $0.69^{+0.1}_{-0.13}$ & 12.0 \\ 
27& $\mathrm{GW190514\_065416}$ & $41.4^{+18.8}_{-9.6}$ & $28.9^{+9.6}_{-9.7}$ & $29.8^{+7.6}_{-5.7}$ & $1.4^{+1.3}_{-0.4}$ & $-0.17^{+0.3}_{-0.35}$ & $3704^{+2662}_{-1939}$ & $0.61^{+0.34}_{-0.28}$ & $67.8^{+18.5}_{-12.5}$ & $0.6^{+0.11}_{-0.18}$ & 8.1 \\ 
\rowcolor{Gray}28& $\mathrm{GW190517\_055101}$ & $38.8^{+10.5}_{-8.2}$ & $24.4^{+6.4}_{-5.8}$ & $26.6^{+3.2}_{-3.6}$ & $1.6^{+0.9}_{-0.5}$ & $0.52^{+0.16}_{-0.19}$ & $1839^{+1438}_{-847}$ & $0.34^{+0.21}_{-0.14}$ & $59.6^{+8.1}_{-8.2}$ & $0.85^{+0.04}_{-0.07}$ & 11.4 \\ 
29& $\mathrm{GW190519\_153544}$ & $63.7^{+10.7}_{-11.2}$ & $39.7^{+12.4}_{-13.6}$ & $43.0^{+7.0}_{-7.6}$ & $1.6^{+1.1}_{-0.5}$ & $0.32^{+0.2}_{-0.25}$ & $2669^{+1877}_{-1021}$ & $0.46^{+0.26}_{-0.16}$ & $97.7^{+12.8}_{-12.4}$ & $0.76^{+0.08}_{-0.12}$ & 13.6 \\ 
\rowcolor{Gray}30& $\mathrm{GW190521\_030229}$ & $101.7^{+27.6}_{-21.4}$ & $57.6^{+19.6}_{-28.7}$ & $64.9^{+12.1}_{-14.2}$ & $1.8^{+2.8}_{-0.6}$ & $-0.21^{+0.48}_{-0.4}$ & $2833^{+2705}_{-1412}$ & $0.49^{+0.36}_{-0.22}$ & $152.0^{+25.5}_{-17.0}$ & $0.54^{+0.21}_{-0.47}$ & 15.2 \\ 
31& $\mathrm{GW190521\_074359}$ & $42.7^{+4.4}_{-4.0}$ & $33.2^{+5.9}_{-5.3}$ & $32.6^{+2.8}_{-2.4}$ & $1.3^{+0.3}_{-0.2}$ & $0.09^{+0.1}_{-0.11}$ & $1112^{+410}_{-434}$ & $0.22^{+0.07}_{-0.08}$ & $72.1^{+5.4}_{-4.4}$ & $0.7^{+0.04}_{-0.04}$ & 24.4 \\ 
\rowcolor{Gray}32& $\mathrm{GW190527\_092055}$ & $37.5^{+19.7}_{-9.1}$ & $20.9^{+9.3}_{-8.7}$ & $23.6^{+6.7}_{-3.7}$ & $1.8^{+2.6}_{-0.7}$ & $0.09^{+0.23}_{-0.27}$ & $2340^{+1903}_{-1119}$ & $0.42^{+0.27}_{-0.18}$ & $56.3^{+17.2}_{-8.6}$ & $0.67^{+0.1}_{-0.24}$ & 8.7 \\ 
33& $\mathrm{GW190602\_175927}$ & $70.6^{+16.9}_{-13.4}$ & $43.6^{+14.4}_{-17.3}$ & $47.3^{+8.4}_{-8.7}$ & $1.6^{+1.6}_{-0.5}$ & $0.12^{+0.24}_{-0.27}$ & $2897^{+1713}_{-1145}$ & $0.5^{+0.23}_{-0.17}$ & $108.8^{+15.7}_{-13.8}$ & $0.69^{+0.1}_{-0.16}$ & 12.2 \\ 
\rowcolor{Gray}34& $\mathrm{GW190620\_030421}$ & $63.5^{+24.7}_{-16.2}$ & $28.0^{+14.5}_{-11.6}$ & $35.9^{+7.7}_{-6.9}$ & $2.3^{+2.6}_{-1.1}$ & $0.26^{+0.23}_{-0.35}$ & $2721^{+1538}_{-1228}$ & $0.47^{+0.21}_{-0.19}$ & $88.5^{+19.4}_{-12.9}$ & $0.72^{+0.11}_{-0.32}$ & 12.0 \\ 
35& $\mathrm{GW190630\_185205}$ & $33.7^{+6.5}_{-5.2}$ & $23.0^{+5.1}_{-4.7}$ & $24.0^{+2.3}_{-1.5}$ & $1.5^{+0.7}_{-0.4}$ & $0.12^{+0.14}_{-0.15}$ & $1192^{+498}_{-409}$ & $0.23^{+0.08}_{-0.07}$ & $54.3^{+4.2}_{-3.9}$ & $0.7^{+0.05}_{-0.09}$ & 15.3 \\ 
\rowcolor{Gray}36& $\mathrm{GW190701\_203306}$ & $55.2^{+10.6}_{-7.4}$ & $41.2^{+8.4}_{-11.4}$ & $41.0^{+5.0}_{-4.8}$ & $1.3^{+0.8}_{-0.3}$ & $-0.09^{+0.22}_{-0.28}$ & $2015^{+691}_{-671}$ & $0.37^{+0.1}_{-0.11}$ & $91.9^{+10.0}_{-8.4}$ & $0.64^{+0.08}_{-0.13}$ & 11.8 \\ 
37& $\mathrm{GW190706\_222641}$ & $70.2^{+13.6}_{-15.2}$ & $37.0^{+13.7}_{-13.1}$ & $43.0^{+9.1}_{-7.2}$ & $1.9^{+1.4}_{-0.7}$ & $0.19^{+0.28}_{-0.35}$ & $4111^{+2188}_{-1730}$ & $0.66^{+0.28}_{-0.24}$ & $101.4^{+16.3}_{-12.5}$ & $0.71^{+0.11}_{-0.22}$ & 12.6 \\ 
\rowcolor{Gray}38& $\mathrm{GW190707\_093326}$ & $12.2^{+2.4}_{-2.3}$ & $7.8^{+1.4}_{-1.3}$ & $8.4^{+0.4}_{-0.3}$ & $1.6^{+0.7}_{-0.5}$ & $-0.04^{+0.1}_{-0.09}$ & $901^{+278}_{-332}$ & $0.18^{+0.05}_{-0.06}$ & $19.1^{+1.4}_{-1.2}$ & $0.64^{+0.03}_{-0.03}$ & 12.8 \\ 
39& $\mathrm{GW190708\_232457}$ & $17.8^{+4.5}_{-2.4}$ & $13.0^{+2.0}_{-2.9}$ & $13.2^{+0.7}_{-0.7}$ & $1.4^{+0.8}_{-0.3}$ & $0.01^{+0.12}_{-0.09}$ & $873^{+322}_{-329}$ & $0.18^{+0.06}_{-0.06}$ & $29.5^{+2.4}_{-1.6}$ & $0.67^{+0.03}_{-0.05}$ & 12.7 \\ 
\rowcolor{Gray}40& $\mathrm{GW190719\_215514}$ & $38.0^{+41.5}_{-11.1}$ & $20.7^{+13.8}_{-8.2}$ & $23.6^{+17.1}_{-4.5}$ & $1.8^{+2.7}_{-0.8}$ & $0.22^{+0.36}_{-0.3}$ & $3607^{+3356}_{-1776}$ & $0.6^{+0.42}_{-0.26}$ & $55.9^{+46.0}_{-10.6}$ & $0.72^{+0.13}_{-0.2}$ & 7.9 \\ 
41& $\mathrm{GW190720\_000836}$ & $12.9^{+6.4}_{-2.8}$ & $7.7^{+1.9}_{-2.4}$ & $8.6^{+0.6}_{-0.5}$ & $1.7^{+1.9}_{-0.6}$ & $0.18^{+0.17}_{-0.1}$ & $1055^{+475}_{-413}$ & $0.21^{+0.08}_{-0.08}$ & $19.8^{+4.2}_{-1.9}$ & $0.71^{+0.03}_{-0.04}$ & 10.7 \\ 
\rowcolor{Gray}42& $\textbf{GW190725\_174728}$ & $13.4^{+17.3}_{-4.4}$ & $5.5^{+2.4}_{-2.6}$ & $7.3^{+0.6}_{-0.5}$ & $2.4^{+8.3}_{-1.3}$ & $-0.02^{+0.48}_{-0.35}$ & $1035^{+487}_{-389}$ & $0.2^{+0.08}_{-0.07}$ & $18.3^{+14.9}_{-2.6}$ & $0.58^{+0.13}_{-0.11}$ & 9.5 \\ 
43& $\mathrm{GW190727\_060333}$ & $38.4^{+8.5}_{-5.6}$ & $29.1^{+6.3}_{-8.3}$ & $28.7^{+4.4}_{-3.4}$ & $1.3^{+0.8}_{-0.3}$ & $0.05^{+0.22}_{-0.24}$ & $3102^{+1236}_{-1132}$ & $0.53^{+0.17}_{-0.17}$ & $64.0^{+9.0}_{-6.4}$ & $0.69^{+0.08}_{-0.11}$ & 11.4 \\ 
\rowcolor{Gray}44& $\mathrm{GW190728\_064510}$ & $12.2^{+6.2}_{-2.2}$ & $7.8^{+1.6}_{-2.2}$ & $8.5^{+0.5}_{-0.3}$ & $1.6^{+1.7}_{-0.5}$ & $0.13^{+0.2}_{-0.08}$ & $1000^{+255}_{-368}$ & $0.2^{+0.04}_{-0.07}$ & $19.2^{+4.2}_{-1.1}$ & $0.7^{+0.02}_{-0.04}$ & 12.7 \\ 
45& $\mathrm{GW190731\_140936}$ & $40.9^{+11.9}_{-8.7}$ & $29.4^{+9.6}_{-10.8}$ & $29.5^{+7.1}_{-5.8}$ & $1.4^{+1.2}_{-0.3}$ & $0.02^{+0.26}_{-0.29}$ & $3345^{+2523}_{-1726}$ & $0.56^{+0.33}_{-0.26}$ & $66.7^{+14.4}_{-11.2}$ & $0.68^{+0.09}_{-0.16}$ & 8.1 \\ 
\rowcolor{Gray}46& $\mathrm{GW190803\_022701}$ & $37.6^{+11.1}_{-7.2}$ & $27.5^{+7.8}_{-8.6}$ & $27.5^{+5.7}_{-4.2}$ & $1.3^{+1.0}_{-0.3}$ & $-0.0^{+0.24}_{-0.29}$ & $3289^{+1882}_{-1574}$ & $0.55^{+0.25}_{-0.23}$ & $62.0^{+12.1}_{-8.7}$ & $0.67^{+0.08}_{-0.15}$ & 8.5 \\ 
47& $\mathrm{GW190814\_211039}$ & $23.1^{+1.6}_{-2.4}$ & $2.6^{+0.2}_{-0.1}$ & $6.1^{+0.1}_{-0.1}$ & $8.9^{+1.1}_{-1.5}$ & $-0.01^{+0.07}_{-0.13}$ & $241^{+37}_{-37}$ & $0.05^{+0.01}_{-0.01}$ & $25.4^{+1.5}_{-2.2}$ & $0.27^{+0.03}_{-0.06}$ & 25.1 \\ 
\rowcolor{Gray}48& $\mathrm{GW190828\_063405}$ & $31.8^{+5.2}_{-3.9}$ & $26.6^{+4.9}_{-5.0}$ & $25.0^{+3.7}_{-2.1}$ & $1.2^{+0.4}_{-0.2}$ & $0.18^{+0.15}_{-0.16}$ & $2120^{+719}_{-925}$ & $0.38^{+0.11}_{-0.15}$ & $54.9^{+7.6}_{-4.4}$ & $0.74^{+0.05}_{-0.06}$ & 15.8 \\ 
49& $\mathrm{GW190828\_065509}$ & $23.5^{+6.0}_{-6.0}$ & $10.9^{+3.4}_{-2.2}$ & $13.7^{+1.3}_{-1.0}$ & $2.2^{+1.1}_{-0.9}$ & $0.05^{+0.15}_{-0.16}$ & $1393^{+657}_{-573}$ & $0.27^{+0.11}_{-0.1}$ & $33.2^{+4.6}_{-4.0}$ & $0.63^{+0.06}_{-0.08}$ & 11.1 \\ 
\rowcolor{Gray}50& $\mathrm{GW190910\_112807}$ & $43.5^{+7.7}_{-7.0}$ & $33.5^{+6.4}_{-7.2}$ & $33.0^{+3.9}_{-3.9}$ & $1.3^{+0.5}_{-0.3}$ & $-0.02^{+0.17}_{-0.19}$ & $1609^{+1139}_{-694}$ & $0.3^{+0.17}_{-0.12}$ & $73.5^{+8.1}_{-8.3}$ & $0.67^{+0.06}_{-0.08}$ & 13.5 \\ 
51& $\mathrm{GW190915\_235702}$ & $31.6^{+6.3}_{-4.2}$ & $25.0^{+4.3}_{-4.6}$ & $24.2^{+2.8}_{-1.9}$ & $1.2^{+0.5}_{-0.2}$ & $-0.03^{+0.17}_{-0.21}$ & $1778^{+642}_{-693}$ & $0.33^{+0.1}_{-0.12}$ & $54.0^{+5.8}_{-4.2}$ & $0.66^{+0.06}_{-0.08}$ & 13.2 \\ 
\rowcolor{Gray}52& $\textbf{GW190916\_200658}$ & $45.7^{+17.0}_{-12.3}$ & $24.0^{+13.2}_{-10.4}$ & $28.0^{+9.0}_{-6.4}$ & $1.8^{+2.2}_{-0.8}$ & $0.15^{+0.33}_{-0.3}$ & $4895^{+2814}_{-2333}$ & $0.77^{+0.34}_{-0.32}$ & $67.4^{+17.2}_{-13.5}$ & $0.69^{+0.14}_{-0.22}$ & 7.5 \\ 
53& $\mathrm{GW190924\_021846}$ & $9.1^{+2.6}_{-2.2}$ & $4.8^{+1.5}_{-0.9}$ & $5.7^{+0.2}_{-0.1}$ & $1.9^{+1.1}_{-0.8}$ & $0.05^{+0.16}_{-0.1}$ & $637^{+156}_{-192}$ & $0.13^{+0.03}_{-0.04}$ & $13.3^{+1.8}_{-0.9}$ & $0.65^{+0.03}_{-0.05}$ & 11.8 \\ 
\rowcolor{Gray}54& $\textbf{GW190925\_232845}$ & $20.2^{+3.9}_{-2.5}$ & $15.6^{+2.1}_{-2.6}$ & $15.4^{+1.0}_{-1.0}$ & $1.3^{+0.5}_{-0.3}$ & $0.05^{+0.13}_{-0.12}$ & $961^{+423}_{-319}$ & $0.19^{+0.07}_{-0.06}$ & $34.2^{+2.5}_{-2.2}$ & $0.69^{+0.05}_{-0.05}$ & 9.6 \\ 
55& $\textbf{GW190926\_050336}$ & $40.1^{+19.1}_{-10.4}$ & $23.4^{+10.8}_{-9.2}$ & $25.9^{+9.3}_{-5.5}$ & $1.7^{+1.7}_{-0.6}$ & $-0.04^{+0.26}_{-0.35}$ & $3634^{+3172}_{-1869}$ & $0.6^{+0.4}_{-0.27}$ & $60.9^{+22.9}_{-11.7}$ & $0.63^{+0.11}_{-0.22}$ & 8.5 \\ 
\rowcolor{Gray}56& $\mathrm{GW190929\_012149}$ & $65.5^{+14.4}_{-16.2}$ & $26.4^{+15.8}_{-10.0}$ & $35.1^{+9.6}_{-7.1}$ & $2.5^{+2.1}_{-1.2}$ & $-0.03^{+0.23}_{-0.27}$ & $3114^{+2486}_{-1366}$ & $0.53^{+0.33}_{-0.2}$ & $89.4^{+16.6}_{-14.3}$ & $0.57^{+0.15}_{-0.25}$ & 9.9 \\ 
57& $\mathrm{GW190930\_133541}$ & $11.9^{+5.5}_{-2.0}$ & $8.1^{+1.6}_{-2.3}$ & $8.5^{+0.5}_{-0.4}$ & $1.5^{+1.5}_{-0.4}$ & $0.14^{+0.19}_{-0.13}$ & $772^{+334}_{-317}$ & $0.16^{+0.06}_{-0.06}$ & $19.2^{+3.3}_{-1.3}$ & $0.71^{+0.04}_{-0.05}$ & 9.9 \\ 
     \hline
\end{longtable*}

\subsection{Binary Black Holes}

\begin{figure*}[t]
  \centering
 \hspace*{-1.0cm}
\includegraphics[width=2\columnwidth]{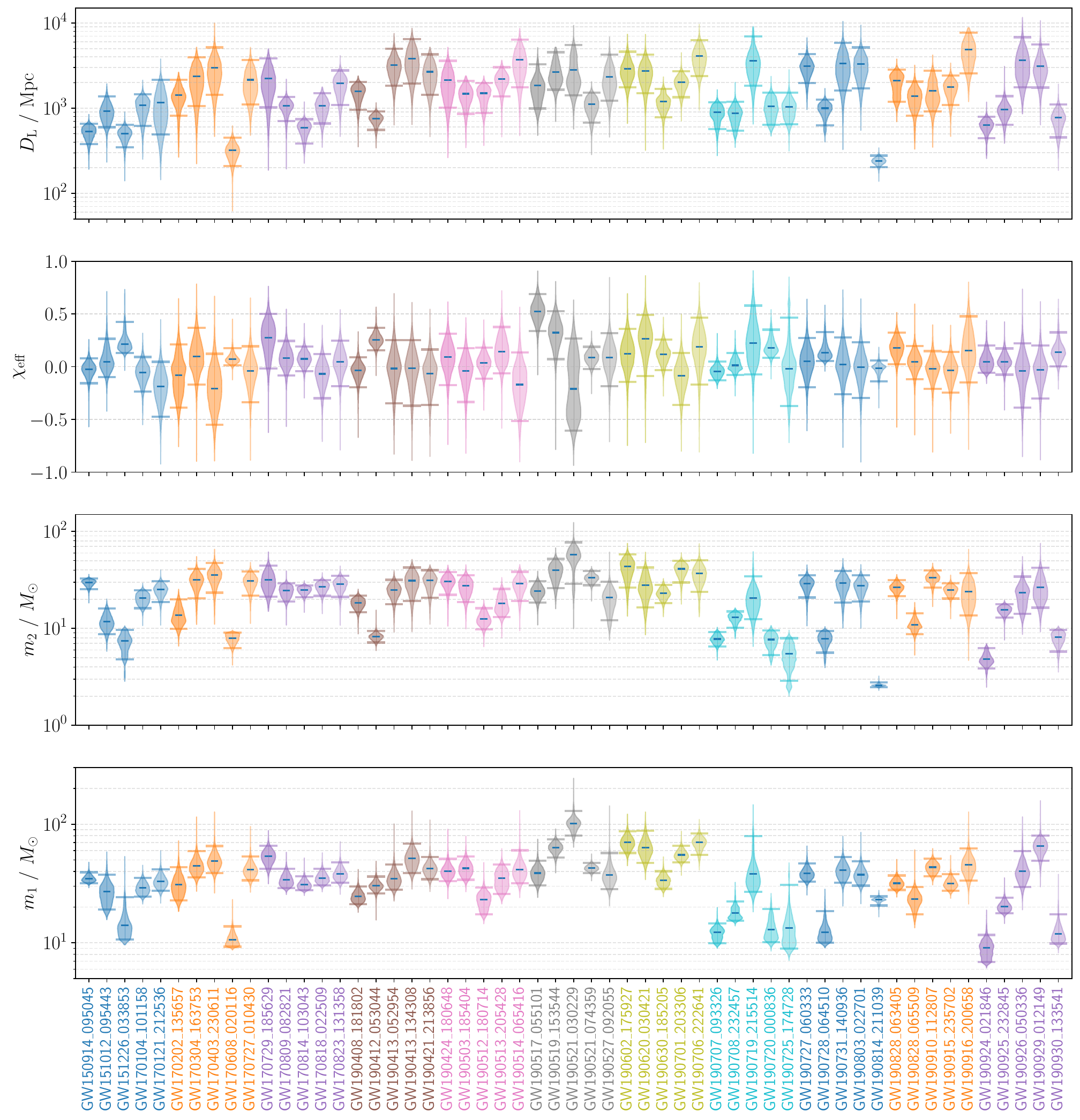}
\caption{The marginalized distributions for component masses $m_1$, $m_2$, the effective spin $\chi_\mathrm{eff}$ and the luminosity distance $D_\mathrm{L}$ for all BBH events detected in 3-OGC.
The median value, the 5th and 95th quantile values are marked with a bar, respectively.
Different colors are used to aid associating each event with its posterior estimates.}
\label{fig:PE-errorbar}
\end{figure*}

\begin{figure*}[t]
  \centering
    \hspace*{-1.8cm}\includegraphics[width=2.6\columnwidth]{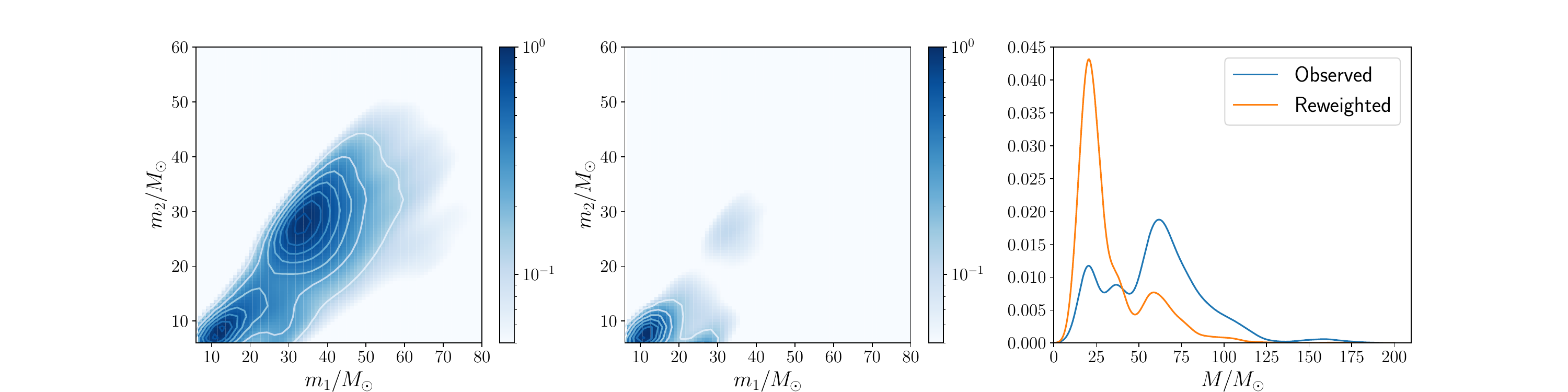}
\caption{Distribution of the source-frame masses of the BBH population from the posteriors obtained from parameter estimation run on all detected BBH events. Here we show detected component mass distribution (left), component mass distribution corrected for the zeroth order selection effect (middle), and the one-dimensional marginals of the total mass distribution (right). The middle plot assumes a constant detection threshold and corrects the distribution for the effect of the signal loudness varying with component mass.}
\label{fig:bbh_population}
\end{figure*}

The mass and spin distributions of the observed population of gravitational-wave mergers, along with their localization posteriors, can be used to constrain various formation channels or population synthesis models \citep{Oshaughnessy:2008,Stevenson:2015,Zevin:2020gbd} and to estimate the rate of mergers~\citep{Roulet:2020wyq, GWTC2-rate}. In Fig.~\ref{fig:PE-errorbar} we show the one-dimensional marginal posteriors on the component masses, effective spin, and luminosity distance for our observed BBH population. Fig.~\ref{fig:bbh_population} shows the combined posterior for all our observed BBH sources, with and without accounting for the zeroth order selection effect introduced by the variation of signal loudness as a function of intrinsic source parameters.

To estimate the source population, we combine the posterior samples for the component masses from each event, to obtain one large collection of mass samples. We do not make additional assumptions about the mass prior and redshift distribution of the population apart from the priors used in parameter estimation. To account for signal loudness, we assign a weight to each sample in the combined posterior that is inversely proportional to the comoving volume that corresponds to the ``horizon distance'' of the given sample. The horizon distance is defined as the maximum luminosity distance an optimally oriented source can be detected with a single-detector threshold SNR of 8.

\vspace{44pt}

\subsubsection{GW190521}

GW190521\_030229 (GW190521) is the most massive confident detection in our catalog. Initial parameter estimates produced by the LIGO and Virgo Collaborations indicated that its component masses were $85^{+21}_{-14}\msun$ and $66^{+17}_{-18}\msun$~\citep{Abbott:2020tfl,Abbott:2020mjq}. This would put at least one of the objects in the ``upper mass gap'' caused by pair-instability supernovae (PISN)~\citep{Woosley:2016hmi,Marchant:2018kun,Stevenson:2019rcw,vanSon:2020zbk}, suggesting that the event may have been created by a hierarchical merger~\citep{Kimball:2020qyd,Liu:2020gif,Fragione:2020han}. This interpretation was challenged in~\citep{Nitz:2020mga} which found multiple modes in the mass posterior. The additional modes were at larger mass ratio (extending to $q\sim6$ or $q\sim10$, depending on the waveform model used), such that component masses straddled the PISN mass gap. However, the highest mass ratio mode (at $q\sim10$) was found by an earlier version of the IMRPhenomXPHM model. An updated version of the IMRPhenomXPHM model (as used in this work) better accounts for the possibility that the total angular momentum could flip direction, inducing transitional precession. With the corrected version of IMRPhenomXPHM, we no longer find significant support for the mass ratio $q\sim10$, however support for the mode at $q\sim6$ remains. This is consistent with the findings of ~\citep{Estelles:2021jnz}. 

An analysis using ringdown quasi-normal modes performed in~\cite{Capano:2021etf} has shown the more equal-mass scenario, however, may be unlikely. The analysis found strong observational evidence for the presence of the $(lmn) = (330)$ sub-dominant harmonic. That a non-zero amplitude was detected for the $(330)$ quasi-normal mode indicates that GW190521 may not be an equal mass binary. 

An electromagnetic counterpart was detected by the Zwicky Transient Factory that may be from the same source as GW190521~\citep{Graham:2020gwr}. If so, this would suggest that GW190521 occurred in the accretion disk of an active galactic nuclei. ~\cite{Nitz:2020mga} found only marginal support for the event to be in coincidence with the electromagnetic signal, with a log Bayes factor of -4 -- 2.3. Using the updated version of IMRPhenomXPHM gives a log Bayes factor of -3.8 -- 2.5.

\vspace{20pt}
\subsubsection{Other multi-modal events}

In addition to GW190521, we find three other events that show second peaks in the likelihood at more asymmetric mass ratios, GW151226\_033853 (GW151226), GW190620\_030421, and GW190725\_174728. However, the prior (which is uniform in component masses) disfavors the higher mass ratio. In addition, for GW190725\_174728 the asymmetric mass portion of the posterior is correlated with a second peak in effective spin at $\chi_{\mathrm{eff}} \sim 0.5$, which is also disfavored by assuming a spin prior that is isotropic in orientation. The combination of the prior and the lower SNR of GW190620\_030421 and GW190725\_174728 results in a weak multimodal structure in the component mass marginal posterior that is less pronounced than it is for GW190521.

A large uncertainty in the mass ratio of GW151226 was found by \citep{mateulucena2021adding} using the same waveform model. More recently, a bimodal distribution in the masses of GW151226 was reported by~\citep{chia2021boxing}, again using the same waveform model. However, \citep{chia2021boxing} found larger support at more asymmetric masses than we do, as well as a secondary peak in chirp mass for which we find weak support. Determining whether these events are truly larger mass ratio than previously expected, or if these secondary modes are due to systematic errors in waveform modelling, will require more study.

\subsubsection{High Mass Ratio Mergers}
The events with the largest (unambiguous) mass ratio are GW190814\_211039 (GW190814) and GW190412\_053044 (GW190412), with a mass ratio of $m_1/m_2 = 8.9^{+1.1}_{-1.5}$ and $3.7^{+1.4}_{-0.9}$, respectively. These estimates are consistent with those found by the LIGO and Virgo Collaborations~\citep{LIGOScientific:2020stg,Abbott:2020khf}. The smaller object in GW190814 had a mass of $2.6^{+0.2}_{-0.1}\,\mathrm{M}_\odot$, making it either the least massive black hole or the most massive neutron star ever detected. If it is a neutron star, it should have a non-zero (albeit small) tidal deformability. Unfortunately, given the high mass ratio and the low signal-to-noise ratio, the event can not bound the tidal deformability away from zero \citep{Flanagan:2007ix}, making it ambiguous whether the object was a neutron star or a black hole~\citep{Abbott:2020khf}. These two events were also the first to have measurable power in sub-dominant harmonics, the $(l,m) = (3,3)$ mode for both~\citep{LIGOScientific:2020stg,Abbott:2020khf}, which can be used to test general relativity as in~\cite{Capano:2020dix}.

\subsection{Neutron Star Binaries}
The only observed neutron star binaries remain the previously reported GW170817~\citep{TheLIGOScientific:2017qsa} and GW190425~\citep{Abbott:2020uma}. The latter is observed in only the LIGO Livingston data, but given its separation from background, and the long duration of the signal which increases the power of signal consistency tests~\citep{Usman:2015kfa}, we consider this detection robust. We obtain a slightly higher estimate for the effective spin of GW190425 than what was reported in~\citep{Abbott:2020uma}. This is due to a difference in prior: as stated above, we use a prior uniform in the spin-component aligned with the orbital angular momentum, whereas~\citep{Abbott:2020uma} used a prior on spin that was isotropic in orientation. Reweighting our posterior to a prior istropic in orientation yields the same effective spin as reported in~\citep{Abbott:2020uma}. We find all other parameters of GW190425 and GW170817 to be consistent with the ``low-spin'' prior results reported in \citep{Abbott:2020uma} and \citep{Abbott:2018wiz}.

GW170817 is the only merger unambiguously observed by electromagnetic emission~\citep{GBM:2017lvd}. Due to the possibility of electromagentic emission from neutron star mergers, we encourage the use of sub-threshold BNS and NSBH candidates released with this catalog to investigate correlations with other archival observations and potentially detect faint sources.

\subsection{Sub-threshold Candidates}
In Table~\ref{table:sub} we show the 17 sub-threshold candidates with $\mathcal{P}_\mathrm{astro} > 0.2$ or IFAR $> 0.5$.
Several sub-threshold candidates have been previously identified. In particular, 151205\_195525 was included in the 2-OGC catalog as a near threshold observation; in our updated analysis it is reduced in significance. 170425\_055334 was previously reported in~\cite{Venumadhav:2019lyq}. 
151011\_192749 was reported in 2-OGC as a sub-threshold event. The majority of these sub-threshold candidates are consistent with BBH mergers. However, 170722\_065503 is consistent with a BNS merger. The full data release includes sub-threshold candidates at lower significance throughout the searched parameter space.

From visual inspection of time-frequency representations of the data around these candidates, there are no signs of loud noise transients that could have caused the corresponding triggers. In a few instances, minor excess power can be observed at frequencies between 50-100 Hz, or at lower frequencies. We cannot conclude if any of these minor power signatures correspond to an instrumental noise artefact or to a marginal astrophysical signal.

\section{Data Release}
We provide analysis configurations, metadata and results at \url{https://github.com/gwastro/3-ogc}~\citep{3-OGC}. The files contain $O(10^6)$ sub-threshold candidates along with their time, SNR, and values for various signal consistency tests. Details of the signal consistency tests and how they are to be interpreted are given in~\cite{Nitz:2017lco} and ~\cite{allen:2004gu}. Each candidate event lists the associated false alarm rate and ranking statistic, to assess their significance. For the most significant candidates inside the focused BBH region discussed in section \ref{sec:multi-det-candidates} we provide an estimate of the probability of astrophysical origin $\mathcal{P}_{\textrm{astro}}$.
We also release our Bayesian parameter inference posterior samples for each of the candidates shown in table \ref{table:peresults} along with a selection of sub-threshold candidates. Additional data products and intermediate results may be made available upon request.

\section{Conclusions}

The 3-OGC catalog of gravitational-wave mergers covers the complete observing period from 2015 to 2019 and includes BNS, NSBH, and BBH candidates. For the first time we include candidates observed by a single sensitive detector. 3-OGC contains the most comprehensive set of merger candidates, including a total of 57 gravitational-wave observations in this period. This includes 4 single-detector mergers in addition to 4 BBH mergers reported here for the first time. We find no additional BNS or NSBH detections beyond the previously reported GW170817 and GW190425. Only the first half of the O3 run which concluded in 2020 has been made public. As the data from the latter half of the observing run is not yet released, the catalog here covers only O1, O2, and O3a. We expect the second half of O3, O3b to be released in 6 months, at which point an updated catalog will be produced.

\acknowledgments

 We thank Bruce Allen and Duncan Brown for providing useful comments. We are grateful to the computing team from AEI Hannover for their significant technical support. MC acknowledges funding from the Natural Sciences and Engineering Research Council of Canada (NSERC). This research has made use of data from the Gravitational Wave Open Science Center (https://www.gw-openscience.org), a service of LIGO Laboratory, the LIGO Scientific Collaboration and the Virgo Collaboration. LIGO is funded by the U.S. National Science Foundation. Virgo is funded by the French Centre National de Recherche Scientifique (CNRS), the Italian Istituto Nazionale della Fisica Nucleare (INFN) and the Dutch Nikhef, with contributions by Polish and Hungarian institutes. 
 
\bibliography{references}

\end{CJK*}
\end{document}